\documentclass[conference,10pt]{IEEEtran}
\IEEEoverridecommandlockouts

\usepackage{times}
\usepackage{amsmath,amsfonts}
\usepackage{algorithmic}
\usepackage{graphicx}
\usepackage{lscape}
\usepackage{multirow}
\usepackage{textcomp}
\usepackage{xcolor}
\usepackage{multirow}
\usepackage{tcolorbox}
\usepackage{url}

\usepackage{breakurl}
\usepackage{array}
\usepackage[acronym,shortcuts]{glossaries-extra}
\glssetcategoryattribute{acronym}{nohyper}{true}
\setabbreviationstyle[acronym]{long-short}

\newacronym{iot}{IoT}{Internet of Things}
\newacronym{csa}{CSA}{Connectivity Standards Alliance}
\newacronym{rra}{RRA}{Read Request Action}
\newacronym{rda}{RDA}{Report Data Action}
\newacronym{end}{End}{End-of-sequence/Ack}
\newacronym{wra1}{WRA-1}{Write Request Action}
\newacronym{wra2}{WRA-2}{Write Response Action}
\newacronym{ira1}{IRA-1}{Invoke Request Action}
\newacronym{ira2}{IRA-2}{Invoke Response Action}
\newacronym{sra1}{SRA-1}{Subscribe Request Action}
\newacronym{sra2}{SRA-2}{Status Response Action}
\newacronym{sra3}{SRA-3}{Subscribe Response Action}
\newacronym{tra}{TRA}{Timed Request Action}
\newacronym{pase}{PASE}{Passcode - Authenticated Session Establishment}
\newacronym{case}{CASE}{Certificate - Authenticated Session Establishment}
\newacronym{chip}{CHIP}{ConnectedHomeIP}
\newacronym{oui}{OUI}{Organization Unique Identifier}

\usepackage{todonotes}

\newcolumntype{P}[1]{>{\centering\arraybackslash}p{#1}}

\def\BibTeX{{\rm B\kern-.05em{\sc i\kern-.025em b}\kern-.08em
    T\kern-.1667em\lower.7ex\hbox{E}\kern-.125emX}}

\begin{document}

\title{Things that Matter - Identifying Interactions and IoT Device Types in Encrypted Matter Traffic
}

\author{\IEEEauthorblockN{Kristopher Alex Schlett\IEEEauthorrefmark{1}, Béla Genge\IEEEauthorrefmark{2},
Ioan Padurean\IEEEauthorrefmark{2},
Savio Sciancalepore\IEEEauthorrefmark{1}
}
\\
\IEEEauthorblockA{
\IEEEauthorrefmark{1}Eindhoven University of Technology, Eindhoven, Netherlands;\\
\IEEEauthorrefmark{2}Bitdefender, Targu Mures, Romania;\\
\IEEEauthorblockA{\textit{\{bgenge, ipadurean\}@bitdefender.com} and \textit{\{k.schlett, s.sciancalepore\}@tue.nl}}}
}

\maketitle

\begin{abstract}
\footnote{This is a personal copy of the authors. Not for redistribution. The final version of the paper will be available through the digital library IEEExplore.}
Matter is the latest interoperability standard for the consumer Internet of Things (IoT) market, developed by the Connectivity Standards Alliance (CSA).
Matter's design pays particular attention to security and privacy: it provides validated secure session establishment protocols, and it uses robust security algorithms to encrypt communications between IoT devices and Matter controllers. However, little is known about the extent to which information can still be inferred from encrypted traffic.

This paper investigates whether a passive adversary can identify Matter interactions and device types by analyzing encrypted traffic metadata.
Using datasets collected from both real-world and simulated Matter deployments, we identify repeatable fingerprints based on packet lengths, directions, and packet sequences.
Our results show that Matter interactions can be identified with over 95\% accuracy even under packet loss and delay conditions. Furthermore, device types such as lighting, lock, and plug appliances can be inferred with accuracy exceeding 88\%. These findings reveal a significant privacy risk affecting Matter-enabled deployments regardless of vendor. The CSA acknowledged our findings and indicated that mitigation mechanisms may be considered in future Matter versions.
\end{abstract}


\maketitle


\section{Introduction} 
\label{sec:intro}

The increasing popularity of the \ac{iot} has led to the definition of several communication technologies and protocol stacks for constrained devices, generating fragmentation and scarce interoperability~\cite{hazra2021_csur},~\cite{lee2021_comst}. Such considerations motivated the \ac{csa} to define {\em Matter}, a new standard enabling devices from various manufacturers and technologies to operate together~\cite{matter_coreSpec}. Started in 2022, Matter today involves more than 550 manufacturers and thousands of devices~\cite{matter_membership}.

One of the major selling points of Matter is {\em security-by-design}~\cite{Homey}. Matter devices feature robust immutable identities verifiable through digital certificates~\cite{duttagupta2025}. Moreover, Matter defines secure onboarding procedures leading to the establishment of session-specific keys, used to authenticate and encrypt all packets~\cite{shafqat2024_spw}.  

At the same time, the explicit focus of Matter on security has motivated intense scrutiny from the security community, resulting in the discovery of a few vulnerabilities, some of them promptly fixed throughout new releases of the standard (see Sect.~\ref{sec:background_related}). However, to the best of our knowledge, no existing research investigates the robustness of Matter-compliant communications to side-channel attacks, particularly to encrypted traffic analysis performed by passive actors possibly breaking the security or exploiting the absence of network security of lower-layer technologies (Wi-Fi, Thread). If successful, an adversary could use encrypted traffic analysis to infer which Matter interactions and device types a user has, without even seeing the deployment. This exposes users to targeted attacks and unauthorized data collection.

Although Matter recommends enabling security at lower protocol layers, practical deployments may expose packet metadata due to weak network credentials, leaked keys, or disabled link-layer protection~\cite{bitdefender2023},~\cite{rezaei2026_eurosp}. Prior research demonstrated that encrypted traffic analysis can reveal sensitive information in many IoT systems~\cite{acar2020_wisec}. However, its impact on Matter has not been systematically investigated.
These considerations motivate our Main Research Question (MRQ):
\begin{tcolorbox}[colback=gray!10, colframe=gray!60!black, boxrule=0.5pt, arc=2pt, left=4pt, right=4pt, top=2pt, bottom=2pt]
\textbf{MRQ}: What information can a fully passive attacker infer by analyzing encrypted Matter traffic? 
\end{tcolorbox}

{\bf Contribution.} In this paper, we investigate experimentally the security of the Matter \ac{iot} standard to encrypted traffic analysis performed by a passive eavesdropper. By analyzing extensive simulated and real-world data, we provide manifold contributions:
\begin{itemize}
    \item We identify fingerprints that allow recognition of encrypted Matter interaction types (Read, Write, Invoke) by only analyzing the direction, length, and sequences of encrypted Matter packets.
    \item We derive device-type fingerprints (e.g., lighting, lock and plug appliances) that generalize across vendors.
    \item We validate the attack on simulated and real deployments and show $>95$\% interaction identification accuracy and $\ge$88\% device-type identification accuracy.
\end{itemize}

All real-world data collected for this work, including 5M+ real-world Matter packets, have been released publicly at~\cite{data}, to support the validation and reproducibility of our research and allow the research community to improve our results.

{\bf Roadmap.} This paper is organized as follows. Sect.~\ref{sec:background_related} introduces Matter and reviews related work, Sect.~\ref{sec:obj} discusses our research objectives, Sect.~\ref{sec:scenario_advModel} introduces our scenario and threat model, Sect.~\ref{sec:methodology} presents our methodology, Sect.~\ref{sec:results} shows the results, Sect.~\ref{sec:discussion} discusses the main implications of our findings and, finally, Sect.~\ref{sec:conclusion} concludes the paper.

\section{Background and Related Work} 
\label{sec:background_related}


{\bf Background.} Matter is an application-layer IoT standard operating over IPv6/UDP and commonly using Wi-Fi or Thread at the MAC layer, coordinated by one or more \emph{Matter Controllers}~\cite{Silicon_Labs_Developer_Documentation2}. The protocol supports five interaction types~\cite{matter_coreSpec}. In a \textit{Read} interaction, the initiator requests one or more attributes or event data from the target. In a \textit{Write} interaction, the initiator attempts to change the values of one or more attributes on the target. In an \textit{Invoke} interaction, the initiator attempts to send one or more commands to the target. \textit{Subscribe} interactions allow the initiator to create subscriptions to attributes or event data, so obtaining periodic updates from the target. Instead, \textit{Report} interactions allow the initiator to report updates to the target.
Finally, regarding network security, Matter devices authenticate each other through PASE or CASE, and derive session keys used to protect communications. Such keys are used for authenticating the Matter header and payload using AES-CCM, and by the default only the Matter payload is encrypted~\cite{Nordic_Semiconductor_2025a}.

{\bf Related Work.} The explicit focus of Matter on security has sparked significant interest by the security research community. 
Duttagupta et al.~\cite{duttagupta2025} recently investigated formally the network security guarantees of Matter. Although its foundations are theoretically sound, they identified certain design choices and implementation flaws introducing vulnerabilities. 
Wang et al.~\cite{wang2025_ndss} identified design flaws in the Matter control capabilities and interfaces exposed to users, which allowed software developers to unintentionally introduce vulnerabilities. 
Ma et al.~\cite{ma2024_usenixSec} used a Large Language Model (LLM) to develop a specification-based fuzzer for Matter, which allowed them to discover various vulnerabilities and risks. They further extended their findings in~\cite{ma2025_mobicom}, and developed an automated system to identify bugs in Matter Software Development Kits (SDKs) that violate the specification.
Shafqat et al.~\cite{shafqat2024_spw} discovered a flaw within the manufacturer’s implementation of the device commissioning process, allowing the adversary to gain unauthorized access. 
Fang et al.~\cite{fang2025_aicc} discovered a design flaw potentially enabling the control and monitoring of the capabilities of Matter devices for IoT users.
Liao et al.~\cite{liao2024_sdiotsec} studied the pairing process of Matter and its delegation phase, revealing the possibility of unauthorized delegation.
Moreover, Genge et al.~\cite{genge_blackEU2024} presented two attacks on Matter devices, namely DeeDoS and Feature Scan, enabling denial of service and unauthorized feature discovery. 
All such works demonstrate that, although considered by design, security in Matter is not perfect. However, none of them investigates to what extent information exposed via encrypted network traffic analysis affects the privacy of Matter users. 
Finally, we acknowledge that many works in recent years have performed encrypted traffic analysis to various IoT protocols~\cite{acar2020_wisec},~\cite{rasool_IoTJ2025},~\cite{msadek_wcnc2019},~\cite{zhang_ccs2018},~\cite{saltaformaggio2016_woot}.
However, none of these works consider Matter-compliant devices, because of their very recent commercialization. We fill this gap through the analysis in this manuscript.

\section{Research Objectives and Motivation} 
\label{sec:obj}

As summarized in the introduction, this paper aims to characterize what information a fully-passive attacker can infer by analyzing encrypted Matter traffic. We break down our overarching MRQ into two separate sub-questions.
We first investigate the presence of recurrent patterns in the encrypted Matter traffic.
This is captured by RQ1:
\begin{tcolorbox}[colback=gray!10, colframe=gray!60!black, boxrule=0.5pt, arc=2pt, left=4pt, right=4pt, top=2pt, bottom=2pt]
\textbf{RQ1:} How accurately can a passive attacker identify specific packets, transactions, and types of interactions from the analysis of encrypted Matter traffic?
\end{tcolorbox}

Existing patterns in encrypted Matter traffic and interactions among devices could represent an effective way for an attacker to monitor the environment without authorization and even without seeing the deployment or knowing which devices are operated. Such monitoring could lead to further physical damage. For example, the existence of a pattern in network traffic indicating the absence of people in a room could inform an intruder about unattended places in homes or offices, serving as the best physical entry points into the area. Therefore, it is important to characterize how accurately a passive attacker can identify interaction types by only using metadata of encrypted Matter traffic. 

Moreover, we investigate to what extent such patterns (RQ1) can be related to specific types of IoT devices in the monitored environment. This is captured by RQ2:
\begin{tcolorbox}[colback=gray!10, colframe=gray!60!black, boxrule=0.5pt, arc=2pt, left=4pt, right=4pt, top=2pt, bottom=2pt]
\textbf{RQ2:} How accurately can a passive attacker infer the usage of specific device types from the analysis of encrypted Matter traffic?
\end{tcolorbox}
The presence of specific device types could provide valuable information for a passive attacker to characterize the digital threat surface of the environment. For example, inferring that access to an area is regulated through a smart lock, an attacker could better focus efforts to get unauthorized physical access by exploiting the presence of such a device (e.g., by performing attacks targeted to that specific device type). Moreover, this information could also be used for unauthorized profiling of users, leading to the exposure of personal preferences and, thus, privacy and regulatory breaches.

Overall, investigating to what extent encrypted Matter traffic could leak information about interactions and device types could inform users about their exposure when using such devices at home or office, better preparing for mitigation.

\section{Scenario and Threat Model} 
\label{sec:scenario_advModel}


\subsection{System Model} 
\label{sec:sysModel}

We consider an \ac{iot} network located in a private home, an office, or a factory. The network includes several Matter-enabled \ac{iot} devices communicating with a Matter controller over Wi-Fi. We also consider the possible existence of devices using Matter over Thread. However, as is usually the case, they communicate with a multi-protocol gateway, e.g., Wi-Fi/Thread border router, translating the Thread traffic into Wi-Fi for the Matter controller~\cite{multiprotocol_gw}. The purpose of such a network is to provide functionality, automate certain processes, and transmit data from multiple sensors to the end user(s). 

We consider that the network is controlled by the end user.
Thus, the user can issue commands to the devices and create new device automations~\cite{yu2021_sensys}. Moreover, we consider that the communication technology used by each device does not change after commissioning, i.e., devices exclusively run Matter over Wi-Fi or Matter over Thread across the deployment period. Matter over Wi-Fi traffic can be sniffed from either the Wi-Fi access point or from any device (including controller) supporting traffic sniffing / capture functions, e.g., through tools like \texttt{tcpdump}.
Depending on the underlying IPv6 network, devices can assume different roles. Within a Wi-Fi network, devices act solely as end devices that communicate with the Wi-Fi access point. Within a Thread network, devices can operate either as end devices or as routers. 

\subsection{Threat Model}
\label{sec:threatModel}

The adversary considered in this paper is a fully-passive attacker, who only eavesdrops packets on both Wi-Fi and Thread networks. The attacker can capture local Wi-Fi or Thread traffic and gain access to the plaintext link-layer traffic, but not to the encrypted application-layer Matter header and payload. Thus, it either possesses the network decryption keys of the Wi-Fi and/or Thread network (e.g., because users set up weak passwords or they are somehow leaked \cite{bitdefender2023}), or it exploits the absence of security at lower layers of these communication technologies, very typical when energy consumption is at budget~\cite{lim2025_arxiv}. Note that Matter is an application-layer protocol. Thus, proper investigation of its robustness against encrypted traffic analysis requires considering full visibility of packet sizes, timings, and directions, but no access to application payloads, in line with relevant literature~\cite{erinola2023usenix},~\cite{xue2024_usenix}.
Overall, the attacker aims to infer network characteristics and device types by observing encrypted Matter traffic generated during regular network communication. We exclude from the attacker scope infrequent events such as pairing, establishment of encrypted sessions, and subscriptions, since typically they are not initiated by users. 

\section{Methodology} 
\label{sec:methodology}


\subsection{Data Collection}
\label{sec:collection}

We collected several datasets from heterogeneous environments, as described in more details below. 

\begin{figure}[t]
    \centering
    \includegraphics[width=.8\columnwidth]{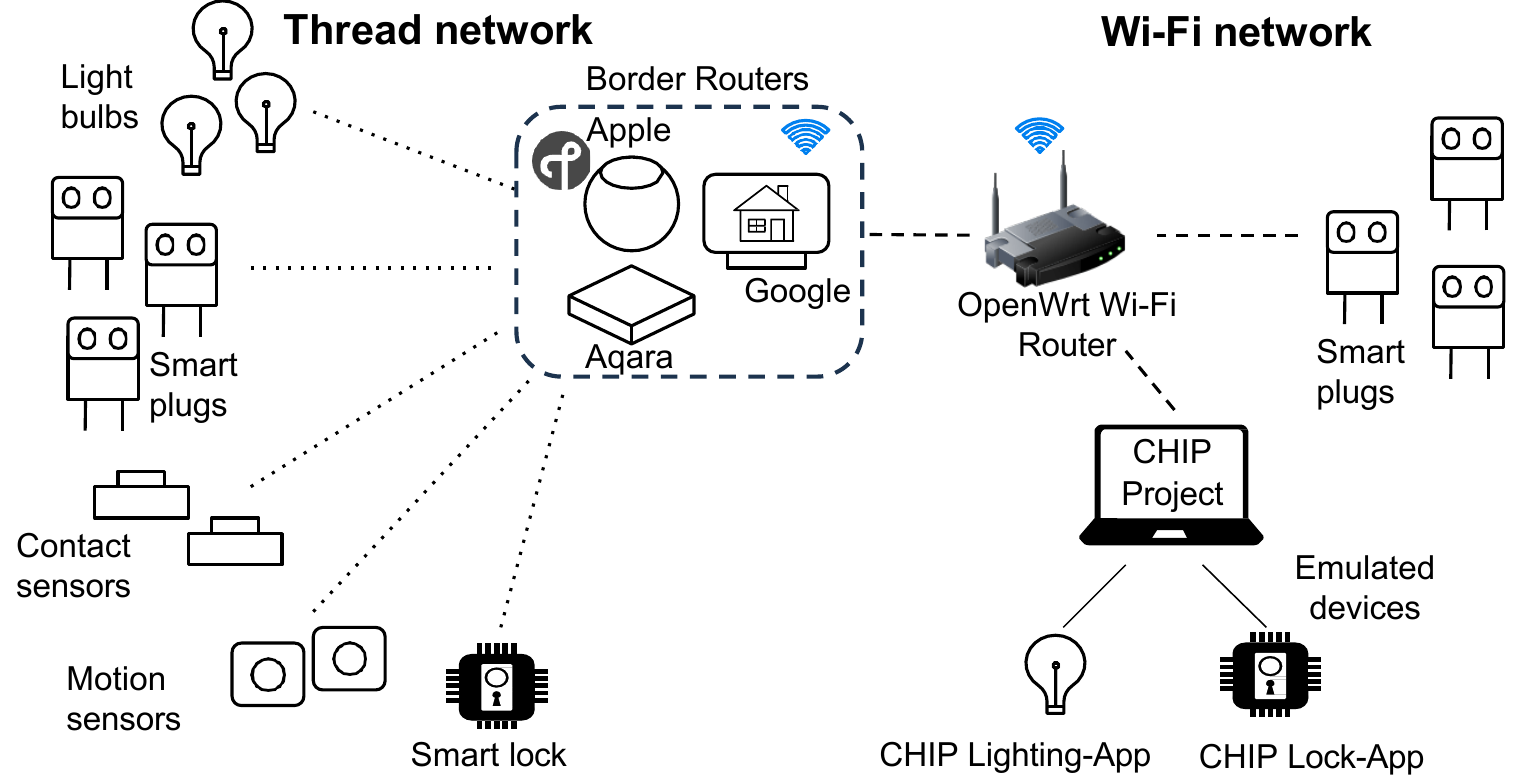}
    \caption{Network configuration of environment E1. 
    }
    \label{fig:e1}
\end{figure}

{\bf E1: Office Environment.} \label{bd-environemnt} This is a real-world Matter network deployed in a real-world office environment, as represented in Fig.~\ref{fig:e1}. 
It includes fourteen (14) Matter-enabled devices of twelve (12) distinct types, including sensors for doors, windows, and motion, as well as plug, lighting, and lock appliances. It also features Matter controllers from various manufacturers. We used two different devices for capturing network traffic, depending on the target protocol. For Matter over Thread, the deployment uses a Nordic nRF52840 DK connected via USB to a laptop running Wireshark with a Nordic NRF extension. For Matter over Wi-Fi, we capture network traffic using \texttt{tcpdump} on an OpenWrt router. We use these passive devices to collect data on the regular behavior of different physical devices communicating via Matter in an office environment.

{\bf E2: Home Environment.} \label{ah-environment}
This environment is a small-scale twin of E1 deployed in an home environment. It includes a smart plug, a smart light bulb, a lock app, a lighting app, and a Home Assistant installation. The applications (i.e., lock app and lighting app) are created from the Matter reference implementation~\cite{Project-Chip}. All devices are in the same fabric and communicate with the Home Assistant controller via Matter. 


{\bf E3: Simulated Environment.} \label{dg-tool}
To analyze Matter data in a controlled environment, we also generated a synthetic dataset by collecting data from a simulated Matter environment used to simulate a Matter fabric. This environment enables troubleshooting, additional investigation of packet structure, and performing attacks without compromising non-simulated devices. We set up three virtual machines running the Ubuntu Server 22.04.5 LTS operating system. On the virtual machines, we downloaded and compiled Matter 1.4 of the \acrshort{chip} GitHub repository by \acrshort{csa} \cite{Project-Chip}, i.e., the reference implementation of Matter for a number of devices, including a controller. 
We reproduced the typical Matter transactions \textit{Invoke, Write, Read}, and \textit{Report}, and collected data with the desired interaction time and packet loss patterns, allowing us to investigate the impact of such factors on the answer(s) to our Research Question(s).  

{\bf Datasets.}  Table~\ref{tab:datasets} summarizes the datasets we collected, available for download at~\cite{data}. 
\begin{table}[t]
\centering
\small
\caption{Datasets collected for answering RQ1 and RQ2. 
}
\label{tab:datasets}
\begin{tabular}{|p{1cm}|p{3.3cm}|p{0.5cm}|p{1.2cm}|p{0.6cm}|}
\hline
\multicolumn{1}{|c|}{\textbf{Dataset}} &
  \multicolumn{1}{c|}{\textbf{Purpose}} &
  \textbf{Env.} &
  \multicolumn{1}{c|}{\textbf{Format}} &
  \textbf{RQ} \\ \hline
$\textbf{D}_0$ &   Observe patterns in encrypted traffic &   E1 &   PCAP, HA Logs &   RQ1, RQ2 \\ \hline
$\textbf{D}_{1}$ & Evaluate impact of delay and packet loss & E3 & PCAP           & RQ1, RQ2 \\ \hline
$\textbf{D}_2$     & Infer interaction and label packets                & E2 & PCAP, HA Logs & RQ1, RQ2 \\ \hline
$\textbf{D}_{3}$ & Device Classification              & E2 & PCAP           & RQ2 \\ \hline
\end{tabular}%
\end{table}

The dataset $\textbf{D}_0$ consists of encrypted Matter traffic (5M+ real-world Matter packets) collected over the course of four weeks in E1, representing regular user-device interactions and automations. It provides a comprehensive collection of device interactions, thus serving as a baseline to investigate the occurrence of patterns within communications involving specific device types.
$\textbf{D}_{1}$ is a synthetic dataset generated in E3, containing Invoke interactions to the \ac{chip} Lighting App. It allows us to study the patterns identified in $\textbf{D}_0$ in a controlled environment. 
$\textbf{D}_2$ consists of encrypted Matter traffic collected over the course of one week in E2, including packets exchanged from reference applications to the end devices deployed in E2. It allows us to examine to what extent encrypted Matter devices can be labeled, and is used to answer both RQ1 and RQ2.
Finally, $\textbf{D}_{3}$ includes data collected by issuing all user commands to the end devices through Home Assistant in E2, and it is used to answer RQ2. 
We provide more details about the methodology used to collect all data in Sects.~\ref{sec:analysis_RQ1} and~\ref{sec:analysis_RQ2}.

\subsection{Data Preprocessing}
\label{sec:preproc}

We preprocessed the data collected in the environments E1, E2, and E3 to transform the raw \texttt{PCAP} files into a data structure suitable for automated data analysis. For each packet, we extracted timing, direction, source/destination identifiers and packet-size-related Matter metadata.
As an additional cleaning step, we also filtered out retransmitted packets. To this aim, for each packet, we check the existence of multiple message copies, i.e., Matter packets from the same source to the same destination containing the same message counter field. If these occur, we keep only the latest packet instance and discard previous ones, as per the Matter specification. 


\subsection{RQ1: Patterns in Encrypted Matter Traffic}
\label{sec:analysis_RQ1}
Associating network patterns to specific interactions defined in the Matter standard requires identifying patterns in the encrypted traffic corresponding to the occurrence of specific interactions, designing rules to detect such patterns, and testing the robustness of the designed rules to various real-world conditions. 

{\bf Interactions Patterns Identification.} We first analyze the traffic in the dataset $\textbf{D}_0$ to identify patterns in {\em sequences} of regular Matter packets related to the occurrence of specific Matter interactions, e.g., \acp{rra}, \acp{wra1}, \acp{wra2}, \acp{ira1} and \acp{ira2}, to name a few. We define a {\em sequence} as $ x\rightarrow y$, where $ \rightarrow $ indicates that within a short time after the packet $x$ is issued, $y$ follows. We empirically set this time to $0.5$ seconds, based on empirical evidence in the observed traffic.
We focus on regular communication patterns. We first identify the controller as the entity most frequently involved in the communication flow with several different end devices. Within such communication patterns, we specifically look for patterns involving header values, packet flows, and packet sizes. 
In most cases, the direction of the packets (from the controller to the end device, or viceversa) and the length of the packets emerge as the minimal amount of information needed to distinguish between the different categories. 
For example, to identify an instance of a packet of type \ac{ira1}, both the packet lengths of the request and the response are required. In some other cases, only one of them is required. For example, when identifying an instance of a packet of type \ac{end}, the packet length is enough. 
We then generate synthetic data to verify the existence of the identified patterns in a controlled scenario, so generating the dataset $\textbf{D}_{1}$. We use the \ac{chip} tool, a Matter controller implementation that allows us to commission a Matter device into the network and communicate with it using Matter messages~\cite{Project-Chip}. As a reference, we generate traffic of invoke interactions related to various commands issued from the controller to a lighting appliance and collect data for a (simulated) time of over 500 days. 
We choose a lighting appliance (a smart bulb) as it contains the most variety of commands. Moreover, lighting appliances share On/Off commands with different device types, allowing us to detect possible misclassifications. We generate data considering two distinct profiles of interactions per day, to represent various scenarios. $EXP_{1}$ represents realistic device usage patterns, i.e., highest probability (0.3) for 3 invoke interactions per day and lower probabilities (0.1) of  higher (5) or lower (1) invoke interactions per day. Instead, $EXP_{2}$ represents random device usage, i.e., a uniform distribution (probability 0.2) of 1 to 5 invoke interactions per day. 

{\bf Detection Rules Design.} To assess which interaction fingerprint best matches a given set of packets, we adopt nearest-neighbor classification in set space~\cite{cunningham2021_csur}, formalized through the scoring metric in Eq.~\ref{eq:s_x}:
\begin{equation}
\label{eq:s_x}
    S_x = |i_d \cap i_x| + \frac{|i_d \cap i_x|}{|i_x|},
\end{equation}
where $S_x$ is the score for interaction type $x$, $i_d$ is the set of observed packet lengths in the network traffic to the end device $d$, and $i_x$ is the set of packet lengths in the fingerprint of $x$. The resulting score is not bound, and we pick the best match by identifying the maximum score across all fingerprints. If ties occur, we use the number of matched sequences to decide which fingerprint best matches. Finally, we treat the data for each day as an independent sample for classification.

{\bf Testing Robustness of Detection.} We label interactions in the datasets using the identified patterns. If packets do not match any of the rules created, they are labeled as \texttt{unknown}, so as to provide a fall-back option for instances where we cannot predict what occurs. Then, we use the created rule set to classify each packet of the encrypted Matter traffic present in $\textbf{D}_2$. 
Similarly to $\textbf{D}_{1}$, we select as the end device a Lighting appliance, since it offers a wider variety of user initiated interactions. The specific interactions and their configurations are selected to cover each of the three interaction types which a user can initiate: \textit{Read}, \textit{Write}, and \textit{Invoke}. 
For \textit{Read} interactions, the \textit{Read} request includes two attributes to increase identification complexity compared to a single-attribute request. Moreover, the controller automatically issues single-attribute \textit{Read} requests at various points in time. The \textit{Write} interaction focuses on one attribute, as the behavior exhibited by the others is comparable. Although we cannot control \textit{Report} transactions, the end device systematically initiates them regularly, either based on a time delta or a state update. In the case of the single-attribute \textit{Read} and \textit{Report} transactions, the quantity of these two is outside our control; however, their logs are included in the dataset. 

To evaluate the extent to which we can infer interactions, we investigate the following metrics: accuracy, recall, precision, and F1 score. 
Finally, we investigate the robustness of our pattern identification technique to changing traffic conditions. To this aim, we evaluate the impact of different levels of packet loss and delays on the robustness of the classification technique.
To test robustness to packet loss, we randomly remove from the dataset $\textbf{D}_2$ a given percentage of packets, and investigate the resulting performance of our classification method. To test robustness to delays, we apply a delay $\delta$ to an increasing amount of packets in the dataset $\textbf{D}_2$, and investigate the resulting performance of our classification method.
We set the delay to $\delta = 0.5$~s, i.e., the maximum delay observed in $\textbf{D}_0$. 
We use these results to quantify the capability of our fingerprints to classify interactions in non-ideal conditions.

\subsection{RQ2: Device Type Classification} 
\label{sec:analysis_RQ2}

Inferring device types from encrypted Matter traffic requires identifying patterns in the encrypted traffic potentially serving as a fingerprint of a particular device type, and then testing the robustness of such a fingerprint.

{\bf Device Types Identification.} We first examine the dataset $\textbf{D}_0$, looking for patterns in the lengths of the packets \ac{ira1} and \ac{tra} sent from the controller to the end devices that could possibly indicate the deployment of specific device types. We focus on these packets as they contain the \textit{Invoke} request with the command payload sent by the controller, or are associated with time sensitive transactions. Based on the identified packet length fingerprints, we establish categories for device classification. These categories include: (i) a set of packet lengths associated to the command (\acrshort{ira1}); and (ii) a set of reoccurring sequences observed in the issued commands/\textit{Timed Transactions}. 
We reuse the definition of {\em sequence} given for RQ1, as well as the same setting of the time threshold to 0.5 seconds (see Sect.~\ref{sec:analysis_RQ1}).
Commands were chosen to categorize the devices. We computed such patterns by analyzing the structure of the packets in the traffic and the logs collected from devices simulated through the \acrshort{chip} tool, and verified their existence in the traffic collected through Home Assistant. For fields likely to vary in size, such as \texttt{Cluster} or \texttt{Attribute}, we inspected their maximum size (in bytes) in the Matter specification or the \acrshort{chip} implementation. For example, for plug appliances, we identified that they all exhibit a length of 59 in \ac{ira1} packets. We did not consider pairing commands, as all devices execute pairing before joining. 

{\bf Testing Robustness of Device Type Classification.} We label instances of \ac{ira1} and \ac{tra} packets in the dataset $\textbf{D}_{0}$. Then, we applied the fingerprints identified in $\textbf{D}_0$ to classify device types in $\textbf{D}_{1}$ and $\textbf{D}_{3}$. 
The procedure for building $\textbf{D}_{1}$ is the same reported for RQ1. The dataset $\textbf{D}_{3}$ includes data collected by issuing all possible user commands for five times through Home Assistant in E2. 
%
%
To assess which fingerprint matches the best, we use the same metric defined for RQ1 in Eq.~\ref{eq:s_x}, considering $S_x$ as the score for device type $x$.
To evaluate the extent to which we can classify devices using encrypted Matter traffic, we consider the same metrics used for RQ1, i.e., accuracy, recall, precision, and F1 score. 

\section{Results} 
\label{sec:results}


\subsection{RQ1: Patterns in Encrypted Matter Traffic}
\label{sec:results_RQ1}

By analyzing the dataset $\textbf{D}_0$, we discover patterns that allow identifying various types of packets in the encrypted Matter traffic. Table~\ref{tab:patterns} summarizes the identified patterns and describes them.
\begin{table*}[t]
\centering
\caption{List of patterns found in the encrypted Matter traffic.}
\label{tab:patterns}
\small
\begin{tabular}{|p{2.4cm}|c|p{7.1cm}|}
\hline
\textbf{Packet Type(s)} & \textbf{Feature(s) Used} & \textbf{Pattern Description}   \\ \hline

\acrshort{rra} + \acrshort{rda} & Length, Direction &
   Multi-attribute \acrshort{rda} response packet to a controller-issued multi-attribute \acrshort{rra} request has length 1.2 times larger than the request. \\ \hline

Single-attribute \acrshort{rra}    & Length, Direction        & Packet has length in the range [51, 58] and it is issued from the controller to the end device.      \\ \hline

Empty \acrshort{rda}    & Length, Direction        & Packet has length 41 and is issued from the end device to the controller without previous request.                     \\ \hline

Non-empty \acrshort{rda}    & Length, Direction  & Packet has length not in \{34, 41, 42\} and is issued from the end device to the controller without previous request.       \\ \hline

\acrshort{end}    & Length                   & Packet has length 34. \\ \hline

\acrshort{wra1} + \acrshort{wra2}          & Length, Direction        & \acrshort{wra2} Response packet to a controller-issued \acrshort{wra1} request has length in range [62, 69]. \\ \hline
  
\acrshort{ira1} + \acrshort{ira2}          & Length, Direction        & \acrshort{ira2} response packet to a controller-issued \acrshort{ira1} request has length in range [67, 74]. \\ \hline

\acrshort{sra2}   & Length                   & Packet has length 42. \\ \hline

\acrshort{tra}    & Length, Direction        & Packet has length in \{38, 39\} and is issued from controller to the end device.                                        \\ \hline
\end{tabular}%
 \end{table*}
Overall, we could identify patterns in the encrypted Matter packets allowing us to classify packets of the following types: multi-attribute \ac{rra}, single-attribute \ac{rra}, empty \ac{rda}, non-empty \ac{rda}, \ac{end}, \ac{wra1}, \ac{wra2}, \ac{ira1}, \ac{ira2}, \ac{sra2}, and \ac{tra}. Some of them could be identified in isolation: for example, all occurrences of packets of type \ac{end} are characterized by a fixed packet length of 34. Some other packets could be identified in pairs: for example, whenever we see a request packet from the controller to the end device, followed by a response packet in the opposite direction with packet length between 62 and 69 bytes, we could match the request (a \ac{wra1} packet) to the response (a \ac{wra2} packet).
We also notice that the identified patterns are not always unique. There is overlap between some of the patterns, specifically between the pairs (\acrshort{ira1}, \acrshort{ira2}) and (\acrshort{wra1}, \acrshort{wra2}). However, our empirical investigation reveals that \acrshort{wra2} packets with length exceeding 66 Bytes are highly uncommon. Similarly, we rarely observe \acrshort{ira2} packets with lengths exceeding 69 Bytes.

Table~\ref{tab:perf_normal} reports the results of applying the detection rules based on the patterns discussed above on the dataset $\textbf{D}_{1}$. We report separately the performance metrics considering all patterns and unusual patterns, i.e. only the patterns \acrshort{ira1}, \acrshort{rra}, and \acrshort{wra1}. These patterns occur less frequently and are more complex. Moreover, based on previous research~\cite{genge_blackEU2024}, they could indicate direct manipulation of the end device, possibly being important for anomaly detection, although this is out of scope. Evaluating detection performances over these three packet types separately allows us to assess whether we can retain performance under different conditions, rather than letting their effect be masked by more frequent and simpler packet types to identify. 

\begin{table}[h]
\centering
\caption{Encrypted packet labeling performance. 
}
\label{tab:perf_normal}
\small
\begin{tabular}{|c|c|c|c|c|}
\hline
\textbf{\begin{tabular}[c]{@{}c@{}}Packet \\ Types\end{tabular}} & \textbf{\begin{tabular}[c]{@{}c@{}}Accuracy \\ $\left[ \% \right]$\end{tabular}} & \textbf{\begin{tabular}[c]{@{}c@{}}Recall \\ $\left[ \% \right]$\end{tabular}} & \textbf{\begin{tabular}[c]{@{}c@{}}Precision \\ $\left[ \% \right]$\end{tabular}} & \textbf{\begin{tabular}[c]{@{}c@{}}F1 Score \\ $\left[ \% \right]$\end{tabular}} \\ \hline
All                                                              & 99.87                                                           & 99.50                                                         & 99.96                                                            & 99.72                                                           \\ \hline
\begin{tabular}[c]{@{}c@{}}IRA-1, RRA, \\ WRA-1\end{tabular}     & 99.60                                                           & 99.34                                                         & 100                                                              & 99.66                                                           \\ \hline
\end{tabular}
\end{table}
Considering all patterns in Table~\ref{tab:patterns}, our methodology achieves accuracy of $99.87\%$, recall of $99.50\%$, precision of $99.96\%$ and F1 score of $99.72\%$. When focusing on the three most unusual packet types, we notice a marginal decline in the investigated metrics, except for precision ($100\%$), indicating no false positives. Overall, in regular network conditions, our methodology achieves strong performance in all labeling instances, indicating the feasibility of inferring specific device types from encrypted Matter traffic.

{\bf Impact of Packet Loss.} We further investigate the robustness of the identified patterns to packet loss. Table~\ref{tab:loss} presents the performance of our methodology under increasing levels of packet loss. 
\begin{table*}[t]
\centering
\caption{Performance with various amount of introduced packet loss.}
\label{tab:loss}
\small
\begin{tabular}{|p{1.3cm}|c|c|c|c|c|}
\hline
\textbf{Packet Loss [\%]}  & \textbf{Packet Types} & \textbf{Accuracy [\%]} & \textbf{Recall [\%]} & \textbf{Precision [\%]} & \textbf{F1 Score [\%]} \\ \hline
\multirow{2}{*}{5}  & All & 98.43 & 96.56 & 99.16 & 97.80 \\ \cline{2-6} 
                    & IRA-1, RRA, WRA-1 & 95.17 & 94.69 & 99.10 & 96.81 \\ \hline
\multirow{2}{*}{10} & All & 96.98 & 93.35 & 98.47 & 95.72 \\ \cline{2-6} 
                    & IRA-1, RRA, WRA-1 & 90.68 & 89.92 & 98.48 & 93.94 \\ \hline
\multirow{2}{*}{25} & All & 92.83 & 84.48 & 96.45 & 89.42 \\ \cline{2-6} 
                    & IRA-1, RRA, WRA-1 & 76.88 & 75.66 & 96.14 & 84.43 \\ \hline
\multirow{2}{*}{50} & All & 86.86 & 69.75 & 93.70 & 77.28 \\ \cline{2-6} 
                    & IRA-1, RRA, WRA-1 & 54.62 & 52.12 & 93.26 & 66.32 \\ \hline
\end{tabular}%
\end{table*}
For relatively low levels of packet loss, our methodology is very robust. Indeed, it performs comparably to the regular conditions reported in Table~\ref{tab:perf_normal}, still maintaining over $94\%$ and $93\%$ in all performance metrics with an average packet loss of $5\%$ and $10\%$, respectively. 
An average packet loss of $25\%$ has a more significant effect on performances. 
Finally, performances drop significantly with $50\%$ packet loss. Thus, as more network instability is introduced, the capability of our methodology to classify packet types, especially the most complex, degrades significantly. We verified that the performance drop is specifically caused by missing response packet information, leading many more packet types to be classified as \texttt{unknown}. However, the relatively high precision of $\approx 93\%$  at $50\%$ packet loss indicates that when packets are classified, they are highly likely to be classified correctly. 

{\bf Impact of Delay in Packet Arrival Time.} Finally, we evaluate the robustness of our pattern identification methodology to increasing packet delays. 
Table~\ref{tab:delay} summarizes the results of our experiments. 
\begin{table*}[t]
\centering
\caption{Performances with added delay of $0.5$~s on increasing subset of the data.}
\label{tab:delay}
\small
\begin{tabular}{|p{1.6cm}|c|c|c|c|c|}
\hline
\textbf{Packets Delayed [\%]} & \textbf{Packet Types} & \textbf{Accuracy [\%]} & \textbf{Recall [\%]} & \textbf{Precision [\%]} & \textbf{F1 Score [\%]} \\ \hline
\multirow{2}{*}{5}  & All & 97.47 & 94.29 & 98.32 & 96.17 \\ \cline{2-6} 
                    & IRA-1, RRA, WRA-1 & 91.88 & 91.06 & 97.71 & 94.17 \\ \hline
\multirow{2}{*}{10} & All & 95.28 & 89.27 & 96.88 & 92.62 \\ \cline{2-6} 
                    & IRA-1, RRA, WRA-1 & 84.42 & 82.97 & 95.40 & 88.57 \\ \hline
\multirow{2}{*}{25} & All & 90.46 & 78.63 & 93.46 & 84.26 \\ \cline{2-6} 
                    & IRA-1, RRA, WRA-1 & 68.93 & 65.95 & 89.92 & 75.52 \\ \hline
\multirow{2}{*}{50} & All & 87.31 & 71.45 & 91.29 & 78.02 \\ \cline{2-6} 
                    & IRA-1, RRA, WRA-1 & 58.80 & 54.93 & 86.67 & 66.37 \\ \hline
\end{tabular}%
\end{table*}
With an average of $5\%$ of the total packets delayed, we can reliably identify patterns, although slightly worse than regular conditions. This can be noticed by observing that the recall in the overall evaluation decreases by $2.27\%$, reaching $94.29\%$. Unusual packets are more affected by the introduced delay, as accuracy and recall drop to $91.88\%$, and $91.06\%$, respectively. 
Performances decrease more significantly when more packets are affected by the delay. However, with $50\%$ of the packets delayed, we notice that performances are generally higher than their counterparts in Table~\ref{tab:loss} for both investigated subsets. Due to the high amounts of packets delayed, both the request and the response could be delayed. In such instances, we could still identify the packets as expected. Instead, with packet loss, correct classification never occurs, as response packets may only be removed. 
In addition, we note that the precision in classifying all packet types decreases to a greater extent than for packet loss, reaching a low of $91.29\%$ and $86.67\%$, respectively. This result occurs since the number of packets remains unchanged and only their relative sequence change. Thus, both the request and the response have higher chances of being mislabeled to a different packet type (not \texttt{unknown}). In summary, the results indicate that as more network instability in the form of packet delays is introduced, the capability of identifying patterns is more sensitive, due to re-ordering. This sensitivity is also reflected in the precision, as it decreases by a greater extent than for packet loss.\\ 
\begin{tcolorbox}[colback=gray!10, colframe=gray!60!black, boxrule=0.5pt, arc=2pt, left=4pt, right=4pt, top=2pt, bottom=2pt]
\textbf{Key Findings:} Matter feature specific packet lengths and sequences, allowing to identify {\em read}, {\em write}, and {\em invoke} interactions with over 99\% accuracy. \\
Performances stay high with low packet loss and delays ($>90\%$ at 10\% loss).
\end{tcolorbox}

\subsection{RQ2: Device Type Classification}
\label{sec:results_RQ2}

We processed the traffic in the datases $D_0$ through the methodology discussed in Sect.~\ref{sec:analysis_RQ2}.

{\bf Fingerprints of Device Types.} We use the patterns previously described for RQ1 and summarized in Table~\ref{tab:patterns} to identify distinct fingerprints indicating the presence of particular types of Matter end devices, i.e., lock appliances, lighting appliances, plug appliances, and sensors. 
As reported in Table~\ref{tab:deviceTypeFingerprints}, each type of Matter end device has a distinct fingerprint given by the network traffic it produces, based on the observed \acrshort{ira1} packet lengths and \acrshort{ira1}/\acrshort{tra} packet sequences, i.e., groups of packets automatically sent altogether by the controller in response to certain other packets.

\begin{table}[h]
\caption{Identified fingerprints of Matter end devices.}
\begin{center}
\small
\resizebox{\columnwidth}{!}{%
\begin{tabular}{|c|cc|}
\hline
\textbf{End Device}         & \multicolumn{2}{c|}{\textbf{Observed Fingerprint}}                                             \\ \cline{2-3} 
\textbf{Type} &
  \multicolumn{1}{c|}{\textit{\textbf{\acrshort{ira1} Packet Lengths}}} &
  \textit{\textbf{\acrshort{ira1} + \acrshort{tra} Sequences}} \\
\textbf{}                   & \multicolumn{1}{c|}{\textbf{(Bytes)}}          & \textbf{(Bytes $\rightarrow$ Bytes)} \\ \hline
\textit{Lighting Appliance} &
  \multicolumn{1}{c|}{ \{59, 70/71, 72, 73, 75\}} &
  \{$(75 \rightarrow 59)$, $(73 \rightarrow 59)$, $(72 \rightarrow 59)$\} \\ \hline
\textit{Lock Appliance} & \multicolumn{1}{c|}{\{64\}} & \{$(39 \rightarrow 64)$, $(38 \rightarrow 64)$\} \\ \hline
\textit{Plug Appliance} & \multicolumn{1}{c|}{\{59\}} & $\emptyset$                   \\ \hline
\textit{Sensor}         & \multicolumn{1}{c|}{$\emptyset$}     & $\emptyset$                   \\ \hline
\end{tabular}%
}
\label{tab:deviceTypeFingerprints}
\end{center}
\end{table}
For end devices under the category of lighting appliances, we further observed that for the same command, i.e., {\em Change Intensity}, the length of the packets varies depending on the controller which issued the command. Specifically, we observe that an instance of the byte value 71 is observed only in packets sent from the Home Assistant controller, whereas we observe an instance of the byte value 70 in the same position of the packet when the same command is issued from the Apple and Google controllers. 
Moreover, we always observe the \acrshort{ira1} packet lengths for packets transmitted by an end device, regardless of the manufacturer. 
For lock appliance, we identified a sequence of byte values $(39 \rightarrow 64)$ in all encrypted packets delivered by such devices. By analyzing possible values of such bytes through the Matter implementation on \acrshort{chip}, we also identified the sequence of byte values $(38 \rightarrow 64)$ as a possible distinct fingerprint characterizing the packets emitted by such devices. 
We also identified other possible sequences (reported in Table~\ref{tab:deviceTypeFingerprints}) allowing to classify all lighting and lock appliances, regardless of the controller and manufacturer. 
For plug appliances, we always observe \ac{ira1} packet length of value $59$. Finally, our analysis did not reveal patterns for \textit{Sensors}.

{\bf Performance of Device Type Detection.} We labelled encrypted network traffic in $D_0$ according to the above rules. Then, we tested the discovered fingerprints on the datasets $D_{1}$ and $D_{3}$. 
Table~\ref{tab:results_RQ2} reports the results of device type identification using the fingerprints in Table~\ref{tab:deviceTypeFingerprints}. 
\begin{table}[t]
\caption{Device identification performance across different experiments.}
\resizebox{\columnwidth}{!}{%
\centering
\small
\begin{tabular}{@{}|c|c|c|c|c|@{}}
\hline
\textbf{Experiment} & \textbf{Accuracy [\%]} & \textbf{Recall [\%]} & \textbf{Precision [\%]} & \textbf{F1 Score [\%]} \\ \hline
$\textbf{D}_1$ - $EXP_{1}$ & 88.00  & 87.98  & 100.00 & 93.60  \\ \hline
$\textbf{D}_1$ - $EXP_{2}$ & 97.60  & 97.59  & 100.00 & 98.78  \\ \hline
$\textbf{D}_3$ & 100.00 & 100.00 & 100.00 & 100.00 \\ \hline
\end{tabular}%
}
\label{tab:results_RQ2}
\end{table}
We can observe that, when tested on $\textbf{D}_3$, our methodology achieves $100\%$ for all four performance metrics. However, the results for $\textbf{D}_1$ reveal a small drop in overall performance, especially when considering accuracy, recall, and F1 score, while precision remains stable. Recall that for generating $\textbf{D}_3$ we issued all possible commands available to the corresponding device types, whereas for $\textbf{D}_1$  we selected the number and type of commands based on probability distributions aimed at representing different user behaviors. We notice a higher drop when considering $\textbf{D}_1$ - $EXP_{1}$, with a decrease in accuracy of $12\%$. For recall and F1 score, we observe reductions of $12.02\%$ and $6.4\%$. For $\textbf{D}_1$ - $EXP_{2}$, the performance drop is only $2.4 \%, 2.41\%$, and $1.22\%$ on the same metrics, respectively. Investigating further the classification errors, the most diffused cause of misclassifications is the common presence of the byte value $59$ in the \acrshort{ira1} packet length of the fingerprints of both lighting and plug appliances, which makes it difficult to distinguish between them if only that \acrshort{ira1} packet length is observed. The performances in $EXP_{2}$ are higher since specific commands have a higher chance of appearing in the network traffic. The behavior of the metrics precision, accuracy, recall, and F1 score indicates that, under the conditions of the experiments, although false positives are not produced, the identification of true positives becomes less effective with less command diversity. 
These findings suggest that the attacker has higher chances when observing many commands issued to the same device. When only a few commands are available, there are greater chances of fingerprint overlap between devices with similar functionalities, leading to misclassifications. 
\begin{tcolorbox}[colback=gray!10, colframe=gray!60!black, boxrule=0.5pt, arc=2pt, left=4pt, right=4pt, top=2pt, bottom=2pt]
\textbf{Key Findings:} Lock, lighting and plug appliances running Matter are characterized by well-defined fingerprints, allowing them to be identified in encrypted Matter traffic with overwhelming accuracy ($\ge 88\%$). \\
Performances improve when the attacker has access to the traffic generated by many commands issued to a device. 
\end{tcolorbox}

\section{Discussion}
\label{sec:discussion}

{\bf Key Insights.} Our analysis of encrypted Matter traffic shows that, although encrypted at the application layer, \textit{Read}, \textit{Write}, \textit{Invoke}, and \textit{Report} transactions remain identifiable, along with all constituent packet types. By Matter's design choice, the packet structure and use of AES-CCM generates consistent and repeatable packet lengths. As shown in Sect.~\ref{sec:results_RQ1}, these patterns expose distinct fingerprints based on the length and direction for each packet type. Therefore, even with encrypted payloads, a passive eavesdropper can still infer device functionality from observable traffic characteristics.
Moreover, all devices of the same type (e.g., lighting, lock, and plug appliances) exhibit repeatable behavior. Although created by different manufacturers, Matter's interoperability-centered design enforces the use of the same clusters for end devices with similar feature sets. Therefore, the same command issued by a controller results in the generation of encrypted traffic with matching length characteristics across all devices of that type. Using these consistent patterns, we fingerprinted devices based on the characteristics identified in Table~\ref{tab:patterns}. Furthermore, because each command generates a distinct packet length per device type, an eavesdropper can map encrypted traffic directly to command names. This mapping enables the construction of a detailed timeline of interactions, where commands can be explicitly identified despite encryption. 
Thus, using only encrypted traffic, an eavesdropper can determine the device type, identify ongoing interactions, and in some cases even infer which type of controller manages the device. These findings represent a significant breach of users' privacy.

{\bf Implications for Research, Industry, and Practice.} Our key findings have broad privacy and security implications. We found that, also for Matter, features extracted from metadata leak valuable information. Thus, Matter encryption alone does not ensure complete privacy. These results are consistent with prior studies (see Sect.~\ref{sec:background_related}) showing that encrypted IoT traffic can still expose user behaviors and device interactions. However, these results are even more impactful for Matter, due to the advertised focus on security and privacy~\cite{matter_coreSpec}. 
Moreover, we observe that Matter’s strong emphasis on interoperability results in deterministic packet lengths. This design choice introduces a privacy leakage that affects all Matter devices collectively, rather than individual ones. Therefore, our methodology and findings apply to all end devices and extend further to any environment in which they are deployed. 

{\bf Mitigations.} Without losing the benefits of interoperability, introducing controlled and synchronized variability in packet structures could help mitigate encrypted traffic analysis, e.g., via randomized packet padding. According to such a methodology, Matter packets could include a variable portion of application-layer payload filled with unused bytes, included on purpose to protect against encrypted traffic analysis. Although effective, this mitigation is well known to come at the cost of higher energy consumption~\cite{papadogiannaki2021_csur}. Thus, future protocol designs should carefully evaluate such mitigation strategies to preserve interoperability while maintaining performance and privacy. 

{\bf Responsible Disclosure.} We notified our findings to the \acrfull{csa} on October 24, 2025. 
The \ac{csa} acknowledged our findings and their general validity. In their initial risk assessment, they categorized the threat described in this paper as {\em medium severity}, as payloads remain encrypted and attackers require access to the local network. However, the extent of information leakage that we demonstrated was beyond their expectations. Thus, the \ac{csa} expressed the willingness to consider implementing and providing support for ad-hoc mitigation strategies, e.g., controlled variability in packet sizes, to reduce the impact of the threat.

{\bf Limitations.} Our dataset includes four device types from three manufacturers. Although this may represent a limited set of Matter device types, we verified the existence of the identified patterns using the official CHIP tool provided by Matter, running the official version of Matter 1.4. Thus, as also acknowledged by the \ac{csa}, our findings apply in general, beyond these devices. 
Due to the scope of this work, we conducted only limited investigation of device error responses. Nevertheless, we identified that \ac{rda} packets containing errors for single-attribute \ac{rra} requests can be regularly detected. This could pave the way for even more robust device detection.
Finally, we consider an attacker with access to the WiFi and Thread MAC-layer keys. This choice aligns with relevant scientific literature on encrypted traffic analysis of application-layer protocols~\cite{erinola2023usenix},~\cite{xue2024_usenix}. 


{\bf Ethical Considerations.} All real-world data collected and analyzed for this research were collected in a controlled laboratory setup, with no harm and neither privacy breaches to people and end users. We also did not make any modifications to the Matter software and the involved devices that would cause breaking of the respective terms of usage.

\section{Conclusions and Future Work} 
\label{sec:conclusion}

In this paper, we investigate the extent to which encrypted traffic compliant with the latest Matter IoT standard can be used by a passive eavesdropper to infer transactions and device types. Our analysis revealed consistent patterns in the encrypted traffic generated by Matter-enabled IoT devices occurring with the execution of specific transactions, e.g., Read, Write, and Invoke. Moreover, we mapped patterns in sequences of transactions to device types, e.g., lighting, lock, and plug appliances, allowing to identify specific device types used in a Matter IoT network. Our findings are not device-specific: any Matter end device is susceptible to traffic analysis. This raises significant concerns about user privacy in any environment that relies on Matter, which become even more relevant if we consider the security-focused design of Matter. 
Overall, our results demonstrate that, without robust countermeasures against traffic analysis, next-generation protocols such as Matter, despite employing strong encryption, risk undermining the very privacy they aim to protect. Future work should focus on extending the analysis to scenarios where the attacker does not know lower-layer keys, and to the use of pattern matching techniques based on Deep Learning.

\section*{Acknowledgements}
This work has been supported by the INTERSECT project, Grant No. NWA.1162.18.301, funded by the Netherlands Organization for Scientific Research (NWO). 
Any opinions, findings, conclusions, or recommendations expressed in this work are those of the author(s) and do not necessarily reflect the views of NWO.

\bibliographystyle{ieeetr}
\bibliography{references}

\end{document}